\documentclass[prx,aps,twocolumn,superscriptaddress]{revtex4-2}

\usepackage[pdftex]{graphicx}
\usepackage{amsbsy,amssymb,amsmath,bm,mathtools}
\usepackage{amsfonts}
\usepackage{mathrsfs}
\usepackage{makecell}
\usepackage{physics}

\usepackage{xspace}
\usepackage{bm}
\usepackage{color}
\usepackage{url}
\usepackage[section]{placeins}
\usepackage[colorlinks=true,linkcolor=blue,citecolor=blue]{hyperref}

\begin{document} 

\title{Gauge-Equivariant Graph Neural Networks for Lattice Gauge Theories}

\author{Ali Rayat}
\affiliation{Department of Physics, University of Virginia, Charlottesville, VA 22904, USA}

\author{Yaohang Li}
\affiliation{Department of Computer Science, Old Dominion University, Norfolk, VA 22904, USA.}

\author{Gia-Wei Chern}
\affiliation{Department of Physics, University of Virginia, Charlottesville, VA 22904, USA}

\date{\today}

\begin{abstract}
Local gauge symmetry underlies fundamental interactions and strongly correlated quantum matter, yet existing machine-learning approaches lack a general, principled framework for learning under site-dependent symmetries, particularly for intrinsically nonlocal observables. Here we introduce a gauge-equivariant graph neural network that embeds non-Abelian symmetry directly into message passing via matrix-valued, gauge-covariant features and symmetry-compatible updates, extending equivariant learning from global to fully local symmetries. In this formulation, message passing implements gauge-covariant transport across the lattice, allowing nonlocal correlations and loop-like structures to emerge naturally from local operations. We validate the approach across pure gauge, gauge–matter, and dynamical regimes, establishing gauge-equivariant message passing as a general paradigm for learning in systems governed by local symmetry.
\end{abstract}

\maketitle

\section{introduction}

\label{sec:intro}

Machine learning models for physical systems increasingly rely on symmetry as a guiding principle. Across quantum chemistry, materials science, and many-body physics, neural networks now serve as efficient surrogates for expensive computations, predicting energies, forces, and observables with near first-principles accuracy while enabling simulations at previously inaccessible scales~\cite{carleo19,sarma19,bedolla21,schwartz21,karniadakis2021,boehnlein22,karagiorgi22,ramprasad2017,butler2018,schmidt2019,bapst2020}. A central insight underlying these advances is that physical laws impose strict transformation rules under symmetry operations, and embedding these constraints directly into model architectures leads to improved data efficiency, stability, and physical consistency.

Equivariant neural networks (ENNs) formalize this idea by enforcing symmetry transformation laws at every layer through representation-theoretic constructions~\cite{cohen2016,cohen2018,weiler2018,kondor2025}. In molecular modeling, for example, $E(3)$-equivariant networks ensure that predicted forces transform consistently under rotations, yielding highly accurate interatomic potentials~\cite{batzner2022,kaba2022,musaelian2023,gong2023,batatia2025,yang2025}. More broadly, equivariance replaces both data augmentation and handcrafted descriptors with exact architectural constraints, establishing symmetry as a central organizing principle in modern scientific machine learning.

Graph neural networks (GNNs) provide a complementary perspective, offering a flexible framework for modeling structured physical systems~\cite{scarselli2009,gilmer2017,hamilton2017,xu2019,maron2019}. By representing systems as graphs—with vertices encoding local degrees of freedom and edges encoding interactions—GNNs naturally incorporate locality, weight sharing, and permutation equivariance. This abstraction is particularly well suited to lattice-based many-body systems, where degrees of freedom may reside on sites, links, or both. In this setting, message passing defines a local propagation of information that, when iterated, captures increasingly extended correlations.

Despite these advances, most existing equivariant architectures are built around \emph{global} symmetry actions, such as rotations, translations, or permutations. Local gauge symmetry, by contrast, defines a fundamentally different structure: physically equivalent configurations are related by independent symmetry transformations at every lattice site. This locality of symmetry cannot be captured by standard equivariant constructions, which are intrinsically tied to a single global group action. Moreover, in non-Abelian gauge systems, physical observables arise from extended Wilson lines and fermion-mediated processes, reflecting intrinsically nonlocal dependencies that are not naturally represented within conventional local architectures.

Lattice gauge theory (LGT) provides the canonical framework for systems governed by local gauge symmetry on a discrete lattice~\cite{wilson1974,kogut1975,fradkin1979,kogut1979,creutz1983,rothe2012}. In high-energy physics, it furnishes a nonperturbative formulation of quantum field theories, with group-valued link variables encoding gauge connections and Wilson loops capturing physical observables. Beyond this setting, gauge structures also emerge in strongly correlated quantum matter, where fractionalized phases are described by emergent gauge fields~\cite{wen2004,lee2006,baskaran1987,affleck1988,wen1991,wen2002,savary2017,zhou2017}. Quantum link models (QLMs) further extend this framework by representing gauge fields with finite-dimensional quantum degrees of freedom that preserve exact local symmetry and admit realizations in atomic, molecular, and optical systems~\cite{chandrasekharan1997,brower1999,wiese2013,banerjee2012,zohar2013,tagliacozzo2013,zohar2016,cheng2024,surace2020,zhou2022}. Together, these examples highlight the broad physical relevance of local gauge symmetry across fundamental, emergent, and engineered systems.

Recent work has begun to incorporate gauge symmetry into machine-learning models for lattice gauge systems~\cite{cranmer2023,tomiya2025}. Gauge-equivariant convolutional architectures construct symmetry-preserving layers by combining parallel transport with bilinear operations, enabling the systematic generation of Wilson loops from local, matrix-valued covariant features~\cite{favoni2022,bulusu2021,lehner2023,lehner2024,apte2024,cohen2019}. In parallel, flow-based and covariant neural-network approaches have been developed for efficient sampling and effective-action modeling, often drawing connections to established constructions such as smearing and gradient flow~\cite{boyda2021,nagai2025,albergo2019,kanwar2020,abbott2022,nagai2023,nagai2024}. 

While these approaches demonstrate that gauge equivariance can be successfully incorporated into neural networks, they remain tied to specific architectural constructions and do not yield a general, scalable representation aligned with the graph structure of lattice gauge systems. In particular, they do not furnish a unified framework capable of handling dynamical matter, fermion-mediated interactions, and intrinsically nonlocal correlations within a single architecture.

To address this, we develop a gauge-equivariant message-passing framework that elevates local symmetry constraints to the level of information propagation. Rather than constructing gauge-invariant features explicitly, the network operates directly on gauge-covariant degrees of freedom, enforcing local symmetry exactly at every layer. In this formulation, learning in lattice gauge systems is recast as the propagation of gauge-covariant information across the lattice, with node and edge features transforming consistently under site-dependent gauge transformations and message updates restricted to symmetry-compatible tensor operations.

Within this framework, iterated message passing effectively implements gauge-covariant transport along lattice paths, allowing Wilson lines and loop-like structures to emerge implicitly from local operations. As a result, intrinsically nonlocal correlations are captured without explicit construction, while global observables can be reconstructed from latent local representations. This unifies local and global structure within a single symmetry-preserving architecture and enables a consistent treatment of static, dynamical, and matter-coupled gauge systems.

More broadly, our results establish gauge-equivariant message passing as a general paradigm for learning in systems governed by local symmetry, extending equivariant machine learning from global group actions to fully local gauge structures and providing a scalable framework for modeling non-Abelian gauge systems across a wide range of physical settings.

\section{Gauge-Equivariant Message Passing}

\label{sec:GNN}

We begin by outlining the essential structure of lattice gauge theory (LGT), a discrete formulation of systems with local gauge symmetry. The defining feature of a gauge theory is local redundancy: physically equivalent configurations are related by independent symmetry transformations at each lattice site. On the lattice, this structure is realized by assigning gauge degrees of freedom to links, rather than to sites or continuous connections. Concretely, group-valued variables $U_{ij} \in \mathcal{G}$ are placed on oriented links $(ij)$, where $\mathcal{G}$ is a compact gauge group. The link variable $U_{ij}$ acts as a discrete parallel transporter from site $i$ to $j$. Under a local gauge transformation,
\begin{equation}
	\label{eq:gauge-U}
	U_{ij} \rightarrow g_i \, U_{ij} \, g_j^\dagger,
\end{equation}
with independent $g_i \in \mathcal{G}$ at each site. Physical observables must therefore be invariant under such transformations and are naturally constructed from closed-loop combinations of links—Wilson loops—which encode gauge flux and characterize confinement, deconfinement, and topological structure. Notably, the placement of degrees of freedom on edges aligns naturally with graph-based representations, where links carry the fundamental variables and vertices define local transformation frames.

In the machine-learning setting, the task is to learn a mapping
\begin{equation}
	\mathcal{F}: \,\, \{U_{ij} \} \, \mapsto\, \mathcal{O}\!\left[ \{U_{ij} \}\right],
\end{equation}
from a gauge-field configuration to physically meaningful observables. The outputs $\mathcal{O}$ may be global quantities—such as total energy or action—or local observables, such as site-resolved densities, currents, or forces. Importantly, these outputs may be either gauge invariant (e.g., energies or Wilson loops) or gauge covariant, transforming consistently under local symmetry operations.

Crucially, the learned mapping must respect the underlying gauge structure. Gauge-equivalent inputs must yield identical invariant outputs, while covariant outputs must transform in accordance with Eq.~(\ref{eq:gauge-U}). This requirement imposes a stringent structural constraint: the model must preserve a site-dependent symmetry action throughout its internal representation. This is particularly important in settings such as quantum link models, where gauge variables correspond to physical degrees of freedom and observables—such as forces or currents—are inherently gauge dependent. Embedding these transformation laws directly into the architecture is therefore not merely a formal constraint, but a guiding design principle that reduces the hypothesis space, improves data efficiency, and enables physically consistent predictions across both invariant and covariant sectors.

\begin{figure*}[t]
\centering
\includegraphics[width=1.99\columnwidth]{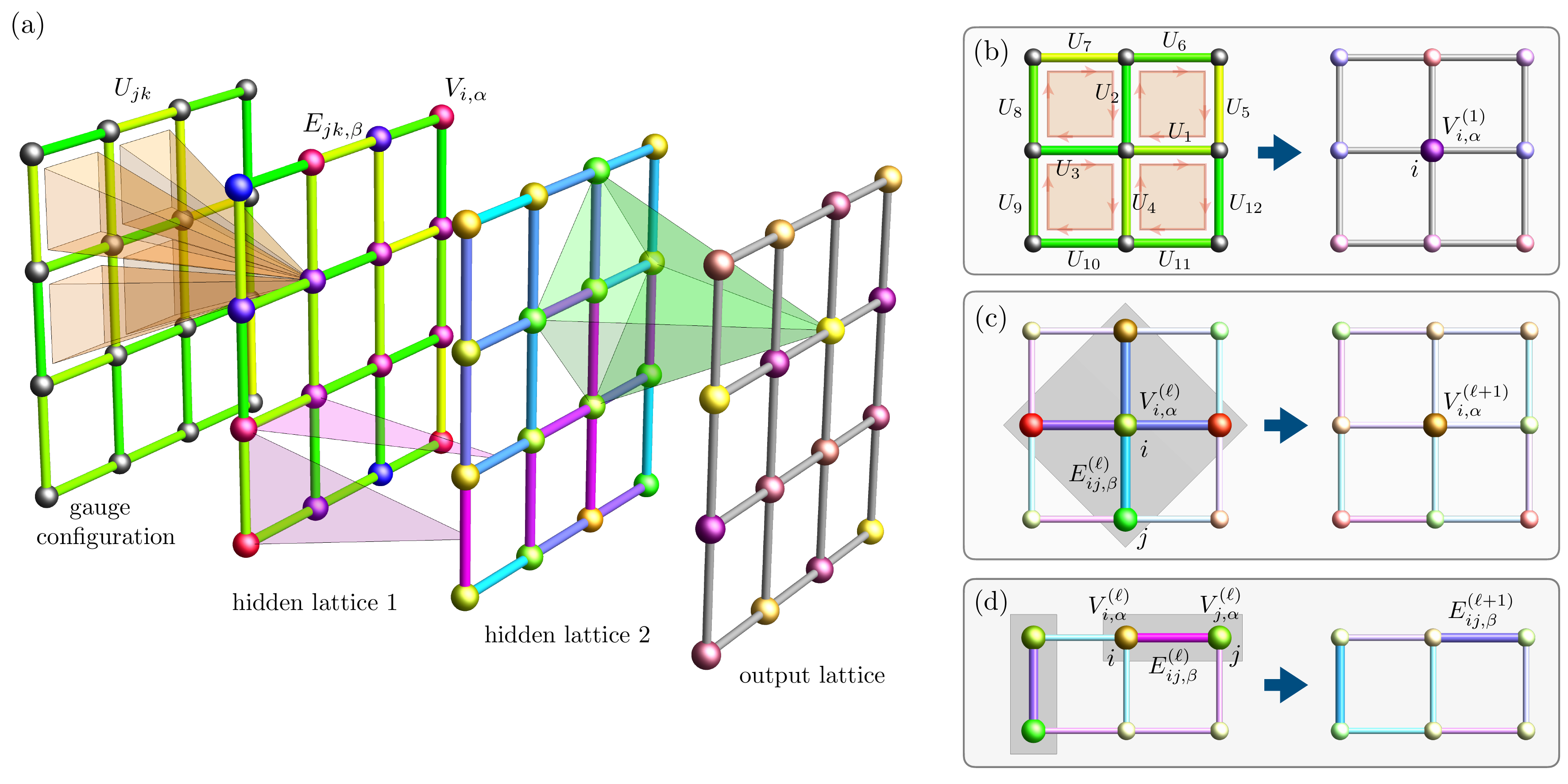}
\caption{Gauge-equivariant GNN architecture. (a) The input consists of matrix-valued gauge links ${U_{jk}}$ defined on the lattice edges. Throughout the network, both node features $V_{i,\alpha}^{(\ell)}$ and edge features $E_{ij,\beta}^{(\ell)}$ are complex $N_f \times N_f$ matrices (elements of $\mathbb{C}^{N_f \times N_f}$), ensuring the correct transformation under local gauge operations. The model is composed of successive hidden lattices followed by an output lattice. (b) Initialization: the first-layer vertex feature $V_{i,\alpha}^{(1)}$ is constructed from ordered products of nearby gauge links $U_{jk}$ around elementary plaquettes adjacent to site $i$, encoding local gauge-covariant information. (c) Node update: at layer $\ell$, the vertex feature $V_{i,\alpha}^{(\ell)}$ is updated via gauge-covariant aggregation of neighboring node and edge features, yielding $V_{i,\alpha}^{(\ell+1)}$ while preserving equivariance. (d) Edge update: the edge feature $E_{ij,\beta}^{(\ell)}$ is likewise mapped to $E_{ij,\beta}^{(\ell+1)}$ through a gauge-covariant transformation involving adjacent vertex features.}
    \label{fig:GNN-scheme2}
\end{figure*}

A natural approach to incorporating gauge symmetry in machine learning is to eliminate gauge redundancy at the level of the input by constructing gauge-invariant features, such as Wilson loops. While this strategy can be effective for Abelian gauge groups—where elementary plaquette fluxes provide a compact and essentially complete local description—it becomes fundamentally limited in the non-Abelian setting. Owing to noncommutativity, Wilson loops are matrix-valued and path-dependent, and no finite set of local loops forms a complete representation of the gauge field. Restricting inputs to small gauge-invariant loops therefore yields an intrinsically incomplete and information-reducing encoding.

More fundamentally, invariant representations discard structured information carried by the link variables themselves. In non-Abelian theories, link matrices $U_{ij} \in \mathrm{SU}(N_f)$ transform covariantly and encode rich internal degrees of freedom that are erased upon reduction to scalar invariants. These considerations motivate a different paradigm: gauge-equivariant architectures that operate directly on link variables and enforce symmetry at the level of the model, retaining full structural information while preserving local covariance.

To preserve gauge covariance throughout the network, we associate both nodes and edges in every layer with matrix-valued features carrying multiple channels indexed by $\alpha, \beta, \cdots$. We group channel-indexed matrices into boldface tensors: $\mathbf V_i = \{ V_{i, \alpha} \}$ and $\mathbf E_{ij} = \{ E_{ij, \alpha} \}$, where each component is an $N_f \times N_f $ matrix. Under a local gauge transformation $\{ g_i \}$, these features transform as
\begin{equation}
	\mathbf V_{i} \to g_i \, \mathbf V_{i} \, g_i^\dagger, \qquad
	\mathbf E_{ij} \to g_i \, \mathbf E_{ij} \, g_j^\dagger,
\end{equation}
where the transformation acts identically on each channel matrix.  This corresponds to adjoint-type transformation for nodes and bi-fundamental transformation for edges. We further impose the Hermiticity constraint $E^\dagger_{ij} = E_{ji}$, such that reversing the orientation of an edge corresponds to Hermitian conjugation.

A schematic of the gauge-equivariant GNN is shown in Fig.~\ref{fig:GNN-scheme2}. For a given gauge configuration $\{U_{ij}\}$, edge features consist of a single channel initialized by the gauge link $\mathbf E^{(1)}_{ij} = U_{ij}$. Node features are constructed from local holonomies to ensure gauge covariance. At each site-$i$, we associate matrix-valued features $V^{(1)}_{i,\alpha}$ given by untraced Wilson loops around elementary plaquettes sharing $i$, with a consistent choice of orientation. 
In addition to these local loops, one may incorporate global or topological information by including Polyakov loops anchored at site $i$, defined along each independent non-contractible direction of the lattice. These provide gauge-covariant probes of the global structure of the gauge field and can be naturally incorporated as additional node features. However, as will be discussed later, such explicit inclusion is not essential for the prediction of gauge-invariant observables. 

Within the message-passing framework, node updates are constructed by aggregating all gauge-covariant contributions that transform consistently at site $i$. These include local interactions among node features as well as edge-mediated contributions from neighboring sites, transported covariantly along lattice links. Concretely, the aggregated message $\mathbf X^{(\ell)}_i = \{ {X}^{(\ell)}_{i,\alpha} \}$ at the $\ell$-th layer takes the form
\begin{eqnarray}
	\mathbf X^{(\ell)}_i &=& \eta_1 \,\mathcal{W}_1\bigl[ \mathbf V_i^{(\ell)} \bigr] 
	+ \eta_2\, \mathcal{W}_2 \bigl[ \mathbf V_i^{(\ell)} \mathbf V_i^{(\ell)} \bigr] \nonumber \\
	&& + \eta_3 \sum_{j \in \mathcal{N}(i)} \mathcal{W}_3\bigl[ \mathbf E^{(\ell)}_{ij} \mathbf V^{(\ell)}_j \mathbf E^{(\ell)}_{ji} \bigr].
\end{eqnarray}
Here, $\mathcal{W}_k[ \cdot ]$ denote learnable linear or multilinear maps acting on channel indices, while matrix multiplication acts on gauge indices and follows the order of factors. The coefficients $\eta_k$ are trainable coefficients controlling the relative contribution of different message channels.

The first two terms are purely local and automatically respect the node transformation law. The linear term performs channel mixing, while the quadratic term introduces local nonlinear interactions through matrix products. Despite their locality, these contributions can encode nontrivial holonomy via the plaquette-based input features. Nonlocal information enters through the edge-mediated term, where neighboring features are parallel transported to site $i$, ensuring consistent transformation under local gauge symmetry.

For notational convenience, we introduce a channelwise normalization operator ${\rm Norm}(\mathbf X) = \left\{ {X_\alpha} / {\lVert X_\alpha \rVert} \right\}$, which normalizes each channel matrix independently. Here $\lVert A \rVert = \sqrt{\mathrm{Tr}(A^\dagger A)}$ denotes the Frobenius norm, which is invariant under gauge transformations. After constructing the message, the updated node feature is given by
\begin{equation}
	\mathbf V^{(\ell+1)}_i = \sigma\Bigl( {\rm Re}\,{\rm Tr}\bigl[ {\rm Norm}\bigl(\mathbf X_i^{(\ell)} \bigr) \bigr] + \mathbf b \Bigr) 
	\odot {\rm Norm}\bigl(\mathbf X_i^{(\ell)} \bigr).
\end{equation}
Here the normalization and nonlinear activation $\sigma(\cdot)$ are applied independently to each channel, $\mathbf b$ is a learnable bias, and $\odot$ denotes elementwise (channelwise) multiplication. Rather than applying nonlinearities directly to matrix elements—which would generally violate covariance—the activation is constructed from a gauge-invariant scalar that modulates the covariant matrix. This scalar gating preserves the transformation structure while enabling expressive nonlinear behavior.

The update of edge features follows the same message-passing principle, with the additional requirement that all contributions transform in the bi-fundamental representation associated with the ordered pair $(i,j)$. The aggregated edge message $\mathbf Y^{(\ell)}_{ij}$ is constructed as
\begin{eqnarray}
	\mathbf Y^{(\ell)}_{ij} &=& \tilde{\eta}_1 \tilde{\mathcal{W}}_1\bigl[ \mathbf E^{(\ell)}_{ij} \bigr]
	+ \tilde{\eta}_2 \tilde{\mathcal{W}}_2\bigl[ \mathbf E^{(\ell)}_{ij} \mathbf V^{(\ell)}_j + \mathbf V^{(\ell)}_i \mathbf E^{(\ell)}_{ij} \bigr] \nonumber \\
	&& + \tilde{\eta}_3 \tilde{\mathcal{W}}_3\bigl[ \mathbf V^{(\ell)}_i \mathbf E^{(\ell)}_{ij} \mathbf V^{(\ell)}_j  \bigr]
\end{eqnarray}
These terms respectively represent channel mixing, minimal coupling to endpoint features, and bilinear dressing from both endpoints, while preserving gauge covariance.

After aggregation, the updated edge feature is then defined as
\begin{equation}
	\mathbf E^{(\ell+1)}_{ij} = \sigma\Bigl( \lVert \mathbf Y^{(\ell)}_{ij} \rVert + \tilde{\mathbf b} \Bigr) \odot {\rm Norm}\bigl(\mathbf Y^{(\ell)}_{ij} \bigr)
\end{equation}
where both the norm $\lVert \cdot \rVert$ and the nonlinear activation are applied independently to each channel matrix.

At the output layer, the form of the prediction is dictated by the transformation properties of the target observable. For gauge-covariant quantities, the output must itself transform covariantly under local gauge rotations. In this case, the network directly outputs node or edge features, $\mathbf V^{(\mathrm{out})}_{i}$ or $\mathbf E^{(\mathrm{out})}_{ij}$, which furnish the appropriate covariant objects.

For gauge-invariant targets, scalar observables are constructed from invariant combinations of the learned features. At the node level, this includes quantities such as $\mathrm{Re}\,\mathrm{Tr}\!\bigl[\mathbf V^{(\mathrm{out})}_{i}\bigr]$, while for edge variables, invariants arise from Wilson-loop constructions, e.g., traces over elementary plaquettes and their higher-order generalizations generated through message passing. 

To incorporate global information, we further include a finite set of Polyakov loops $\mathbf P = \{P_1, \dots, P_D\}$ defined along the independent lattice directions. These invariant features are then processed by a site-wise multilayer perceptron (MLP), shared across all lattice sites,
\begin{equation}
	\mathcal{O}_i = \mathrm{MLP}_{\bm \theta}\Bigl(
	\mathrm{Re}\,\mathrm{Tr}[\mathbf V^{(\mathrm{out})}_{i}],
	\, \mathrm{Re}\,\mathrm{Tr}[\mathbf E^{(\mathrm{out})}_{\Box}],
	\, \mathrm{Re}\,\mathrm{Tr}[\mathbf P]
	\Bigr).
\end{equation}
In this framework, equivariant message passing systematically builds an expressive hierarchy of contractible Wilson loops from local features, whose traces serve as local invariant descriptors. In contrast, non-contractible loops encode global (topological) information and remain insensitive to local equivariant updates. Their contribution thus enters naturally as a finite set of global scalar features at the level of the invariant MLP.

In this way, the gauge-equivariant GNN serves as a structured feature extractor that organizes local and global gauge information into symmetry-compatible representations, with the final MLP mapping these invariant descriptors to physical observables.

\section{Learning Pure Gauge Structure in 3+1 Dimensions}

\label{sec:pure-gauge}

Having established the general framework for gauge-equivariant graph neural networks, we now turn to a concrete benchmark. We consider pure lattice gauge theory in $3+1$ dimensions, where the dynamical degrees of freedom reside solely on the gauge links. This setting provides a nonperturbative discretization of Yang--Mills theory and a minimal testbed for confinement, topology, and non-Abelian gauge dynamics. Physical observables are constructed from gauge-invariant traces of Wilson loops, with elementary plaquettes forming the basic building blocks. As such, closely related models have become standard benchmarks for gauge-equivariant machine learning approaches~\cite{favoni2022}.

This problem offers a particularly transparent probe of inductive bias. The total action and other observables can be expressed as linear combinations of gauge-invariant Wilson loops. If these invariants were provided explicitly, the task would reduce to a simple regression. Here, instead, we ask a sharper question: can a gauge-\emph{equivariant} network discover and organize these invariant structures directly from gauge-covariant inputs, without prior specification of the Wilson-loop basis?

Training and evaluation data are generated using Monte Carlo sampling of $\mathrm{SU}(3)$ lattice gauge theory with the Wilson action,
\begin{equation}
	S[U] \;=\; -\frac{\beta}{2} \sum_{\square} \mathrm{Re}\,\mathrm{Tr}\,
	\Big( U_{ij} U_{jk} U_{kl} U_{li} \Big),
\end{equation}
where the ordered product runs over links around each plaquette $\square$. We focus first on predicting the total action $S[U]$. Rather than constructing it explicitly from plaquettes, we adopt a flexible decomposition $S = \sum_i \epsilon_i$, where $\epsilon_i$ are gauge-invariant scalars produced at the output layer. This Behler--Parrinello--type Ansatz is used here purely as a modeling device: the $\epsilon_i$ are latent variables with no prescribed physical meaning. The key point is that no Wilson-loop structure is hard-coded. This benchmark therefore directly tests whether the network can reconstruct gauge-invariant observables and their additive organization from gauge-covariant representations alone.

As a complementary task, we consider the prediction of the local topological charge density $q_i$, defined on the lattice as
\begin{equation}
q_i \;=\; \frac{1}{32\pi^2} \, \epsilon_{\mu\nu\rho\sigma} \,
\mathrm{Tr}\left[
\frac{U_{i,\mu\nu} - U_{i,\mu\nu}^\dagger}{2i}
\frac{U_{i,\rho\sigma} - U_{i,\rho\sigma}^\dagger}{2i}
\right],
\end{equation}
with the total charge given by $Q = \sum_i q_i$. Here, we treat $q_i$ explicitly as a \emph{local} gauge-invariant observable constructed from nearby plaquettes, rather than emphasizing its global interpretation. As with the action, $q_i$ can be expressed in terms of traces of small Wilson loops and therefore belongs to the same class of structured but ultimately local functionals of the gauge field. In practice, for generic Monte Carlo configurations without smoothing, the summed charge $Q$ is not strictly quantized, further supporting its interpretation here as a continuous observable assembled from local contributions.

From a modeling perspective, this task probes the ability of the network to predict \emph{site-resolved} gauge-invariant quantities, complementing the global action prediction. As in the case of $S[U]$, the underlying functional dependence on Wilson-loop invariants is relatively simple; the key question is whether the equivariant architecture can reconstruct these local invariants directly from gauge-covariant inputs without explicit feature engineering. Together, these tasks provide a controlled setting to assess the expressive power and inductive bias of gauge-equivariant message passing.

\begin{figure}[t]
\centering
\includegraphics[width=0.99\columnwidth]{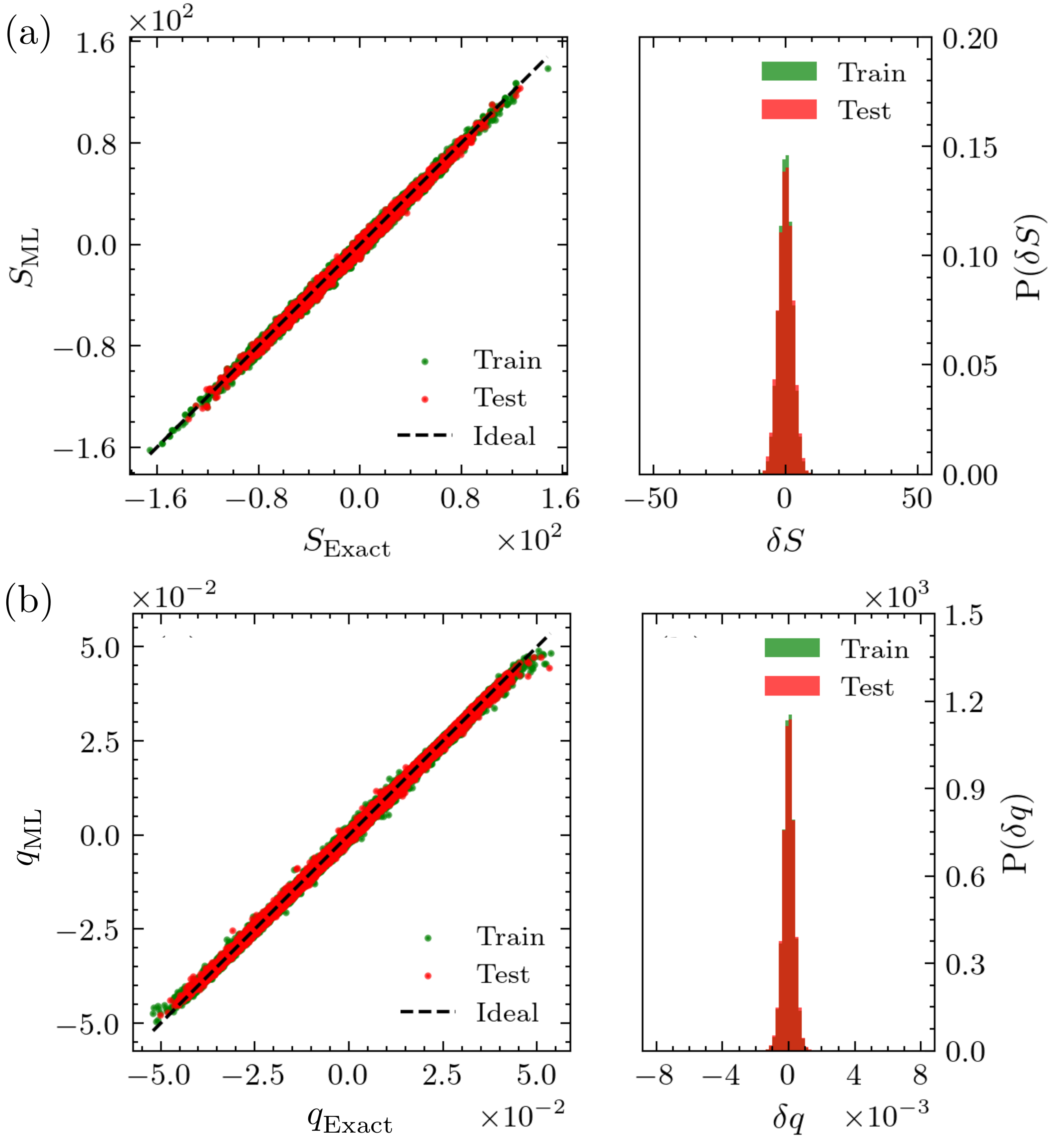}
\caption{
Benchmark of the gauge-equivariant GNN for the $3{+}1$D pure gauge model. (a) Parity plot (left) comparing predicted action $S_{\mathrm{ML}}$ with exact values $S_{\mathrm{Exact}}$ (dashed line: ideal agreement), and error distribution (right) $\delta S = S_{\mathrm{ML}} - S_{\mathrm{Exact}}$ for training (magenta) and test (blue) sets. The model achieves $\mathrm{MSE} = 8.96$ and $R^2 = 0.994$, indicating high accuracy and negligible bias. (b) Same analysis for the local topological charge $Q$, with $\delta Q = Q_{\mathrm{ML}} - Q_{\mathrm{Exact}}$, demonstrating accurate capture of local topological fluctuations (MSE $=1.26\times 10^{-7}$, $R^2 =0.99$).
}
\label{fig:pure-gauge-benchmark}
\end{figure}

The benchmark results are summarized in Fig.~\ref{fig:pure-gauge-benchmark}. Panel (a) shows the prediction of the total action $S[U]$. The parity plot exhibits an essentially perfect linear correlation between predicted and exact values for both training and test sets, with a narrow and symmetric error distribution centered at zero. This indicates that the network accurately reconstructs the additive structure of the action and generalizes well across configurations.
Fig.~\ref{fig:pure-gauge-benchmark}(b) presents the prediction of the local topological charge density $q_i$. Despite the increased difficulty of this site-resolved task, the model maintains strong agreement with the exact values, with only moderate broadening around the ideal diagonal. The corresponding error distribution remains sharply peaked at zero, demonstrating that local fluctuations are captured with high fidelity. Together, these results show that gauge-equivariant message passing accurately recovers both global and local gauge-invariant observables directly from gauge-covariant inputs, without explicit construction of invariant features.

\section{Structure-Property Learning in Gauge-Matter Systems}

\label{sec:gauge-matter}

\begin{figure*}[t]
\centering
\includegraphics[width=1.99\columnwidth]{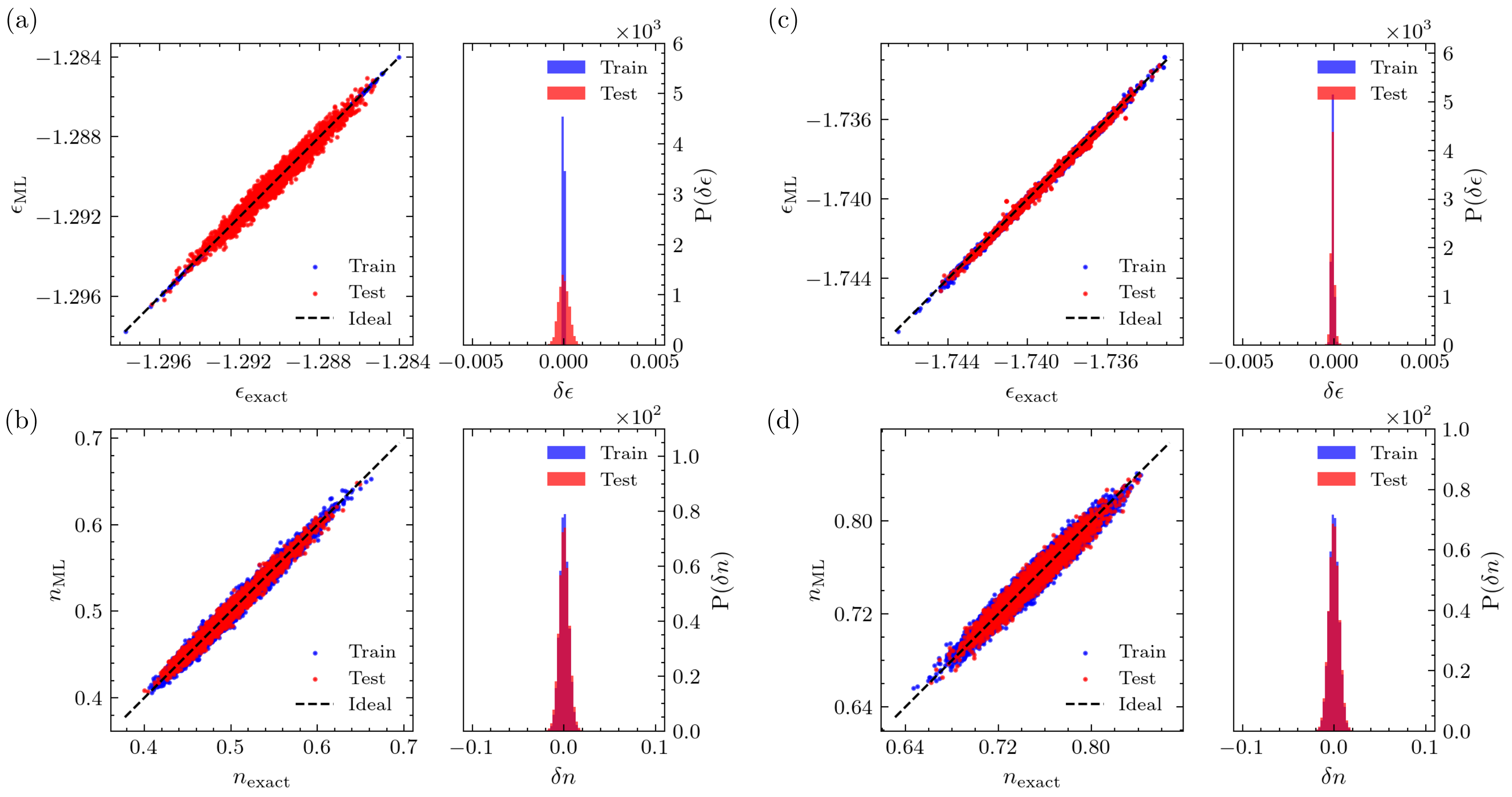}
\caption{Benchmark of the gauge-equivariant GNN for non-Abelian gauge--matter systems on a $20 \times 20$ square lattice. Panels (a,b) correspond to the SU(2) model and panels (c,d) to the SU(3) model. For each case, we report energy density prediction at half filling (a,c) and local fermion density prediction at quarter filling (b,d). Left subpanels show parity plots comparing ML predictions with exact diagonalization (ED) results (dashed line: ideal agreement), while right subpanels display the corresponding error distributions, $\delta \mathcal{O} = \mathcal{O}_{\mathrm{ML}} - \mathcal{O}_{\mathrm{ED}}$, for training (magenta) and test (blue) sets. For SU(2), the energy model achieves $\mathrm{MSE} = 6.713 \times 10^{-8}$ with $R^2 = 0.9732$, and the density model yields $\mathrm{MSE} = 2.8878 \times 10^{-5}$ with $R^2 = 0.9759$. For SU(3), the corresponding values are $\mathrm{MSE} = 1.615 \times 10^{-8}$ with $R^2 = 0.9941$ for energy, and $\mathrm{MSE} = 3.2615 \times 10^{-5}$ with $R^2 = 0.9525$ for density. }
    \label{fig:benchmark-nonabelian}
\end{figure*}

Having established a controlled benchmark in the pure gauge setting, we now turn to a more demanding and physically richer problem: learning gauge--matter coupling. In contrast to pure gauge theory, where observables are constructed directly from local Wilson loops, the presence of dynamical matter fields induces effective interactions between gauge links that are mediated by fermionic degrees of freedom. These interactions are generically nonlocal, reflecting the extended nature of fermionic wavefunctions and their sensitivity to the global gauge background. As a result, predicting physical observables in gauge--matter systems constitutes a substantially more stringent test of the expressive power of gauge-equivariant architectures.

Gauge theories generically include dynamical matter fields transforming nontrivially under $\mathcal{G}$. A minimal lattice realization couples site fermions to gauge links via
\begin{equation}
	\label{eq:LGT-matter}
	\mathcal{H} = \sum_{\langle ij \rangle} \psi_i^\dagger U_{ij} \psi_j + \mathrm{h.c.},
\end{equation}
where $\psi_i$ denotes a fermionic annihilation operator in the fundamental representation, and $U_{ij} \in \mathcal{G}$ is the link variable. Under local gauge transformations $\psi_i \to g_i \psi_i$ and $U_{ij} \to g_i U_{ij} g_j^\dagger$, the Hamiltonian remains invariant by construction. Hermiticity further requires $U_{ji} = U_{ij}^\dagger$, consistent with link orientation.

Beyond high-energy contexts, Hamiltonians of this form arise naturally in condensed-matter systems. In parton constructions of strongly correlated phases, microscopic degrees of freedom are fractionalized into emergent fermions coupled to a dynamical gauge field. The resulting hopping processes are dressed by gauge connections, leading to effective descriptions identical in structure to Eq.~(\ref{eq:LGT-matter}). In this setting, the gauge field encodes the redundancy of the fractionalized representation, while its dynamics govern confinement, deconfinement, and the emergence of exotic phases such as quantum spin liquids.

We benchmark the gauge-equivariant GNN on $\mathrm{SU}(2)$ and $\mathrm{SU}(3)$ lattice gauge theories defined by Eq.~(\ref{eq:LGT-matter}) on a $20 \times 20$ square lattice. Gauge configurations $\{U_{ij}\}$ are generated randomly to probe a wide region of the configuration space without imposing an explicit equilibrium distribution. For each configuration, the fermionic Hamiltonian is solved via exact diagonalization to obtain the corresponding observables. This dataset construction isolates the structure--property mapping from gauge fields to fermionic responses, enabling a direct assessment of the model's capacity to learn the underlying functional dependence. Despite the absence of equilibrium sampling, such broad, unbiased coverage of configuration space provides a stringent and informative benchmark for evaluating both expressivity and generalization of the learned representation.

We consider two supervised tasks: prediction of the total energy $E$ and the site-resolved fermion density $n_i = \langle \psi_i^\dagger \psi_i \rangle$. Remarkably, the same gauge-equivariant architecture accurately captures both explicitly local quantities such as $n_i$ and global quantities such as $E$, the latter reconstructed through a latent decomposition $E = \sum_i \varepsilon_i$ in terms of learned site-resolved contributions. This formulation parallels the Behler--Parrinello-type decomposition used for the gauge action in the pure gauge case, but here acquires a deeper significance. Because the energy depends on fermionic states delocalized across the entire lattice, the learned $\varepsilon_i$ implicitly encode fermion-mediated correlations that are intrinsically nonlocal.

Within this framework, the quantities $\varepsilon_i$ should be viewed as latent variables that provide a symmetry-consistent decomposition of a global observable, rather than physically well-defined local energies. The key point is that both local observables and global quantities are represented within a unified local embedding: the network constructs site-resolved features that simultaneously predict $n_i$ and, when aggregated, recover $E$. In doing so, it implicitly encodes fermion-mediated correlations over extended length scales, enabling a local, gauge-equivariant architecture to approximate observables governed by globally entangled degrees of freedom.

As summarized in Fig.~\ref{fig:benchmark-nonabelian}, the gauge-equivariant architecture achieves high accuracy across both gauge groups. For $\mathrm{SU}(2)$, the energy prediction attains a test MSE of $6.7\times10^{-8}$ ($R^2=0.973$), while the density task yields $2.9\times10^{-5}$ ($R^2=0.976$). For $\mathrm{SU}(3)$, the energy accuracy further improves to $1.6\times10^{-8}$ ($R^2=0.994$), with density prediction at $3.3\times10^{-5}$ ($R^2=0.953$). Parity plots exhibit a tight collapse onto the diagonal with sharply peaked error distributions, and the close agreement between training and test metrics indicates robust generalization.

Notably, near-saturation accuracy is achieved with only shallow architectures. This does not imply that the learned representation is short-ranged. While message passing is local at the architectural level, the effective interactions are mediated by fermions and are therefore intrinsically nonlocal. In practice, this nonlocality is captured through repeated gauge-covariant transport and contraction operations, which concatenate link variables along extended paths. As a result, even shallow networks generate features corresponding to Wilson lines of increasing spatial extent, and the effective receptive field—measured in gauge transport length—grows rapidly beyond the naive graph distance. This mechanism enables the model to capture fermion-mediated correlations across the lattice despite its local construction.

Taken together, these results demonstrate that gauge-equivariant graph neural networks provide a unified and scalable framework that simultaneously resolves local observables and reconstructs global quantities through latent local representations, even in systems where the underlying physics is governed by intrinsically nonlocal, fermion-mediated interactions.

\section{Learning Gauge Dynamics with Equivariant Force Fields}

\label{sec:gauge-dynamics}

We next demonstrate that gauge-equivariant GNNs provide a natural and efficient framework for learning force fields in dynamical lattice gauge systems, extending beyond the static prediction of observables. As a concrete setting, we consider the $\mathrm{SU}(2)$ quantum link model (QLM), in which gauge fields reside on lattice links as finite-dimensional degrees of freedom and transform covariantly under local $\mathrm{SU}(2)$ rotations [Eq.~(\ref{eq:gauge-U})]. QLMs establish a direct connection between high-energy lattice gauge theories and experimentally accessible platforms, particularly in ultracold-atom realizations where gauge and matter fields emerge from constrained dynamics in optical lattices. Although realistic implementations involve additional constraints and interactions, they retain the essential structure of dynamical gauge--matter coupling.

As a proof of principle, we adopt a minimal QLM-inspired Hamiltonian,
\begin{equation}
	H = \sum_{\langle ij \rangle} \frac{g^2}{2} \, \mathbf{L}_{ij}^2 
	+ \sum_{\langle ij \rangle} \psi_i^\dagger U_{ij} \psi_j ,
\end{equation}
where $U_{ij} \in \mathrm{SU}(2)$ are link variables and $\mathbf{L}_{ij} \in \mathfrak{su}(2)$ are their conjugate generators. In the semiclassical limit, the dynamics is governed by a Lie--Poisson structure inherited from the underlying quantum commutation relations, with $\mathbf{L}_{ij}$ generating left rotations of $U_{ij}$ on the group manifold.

\begin{figure}[t]
\centering
\includegraphics[width=0.99\columnwidth]{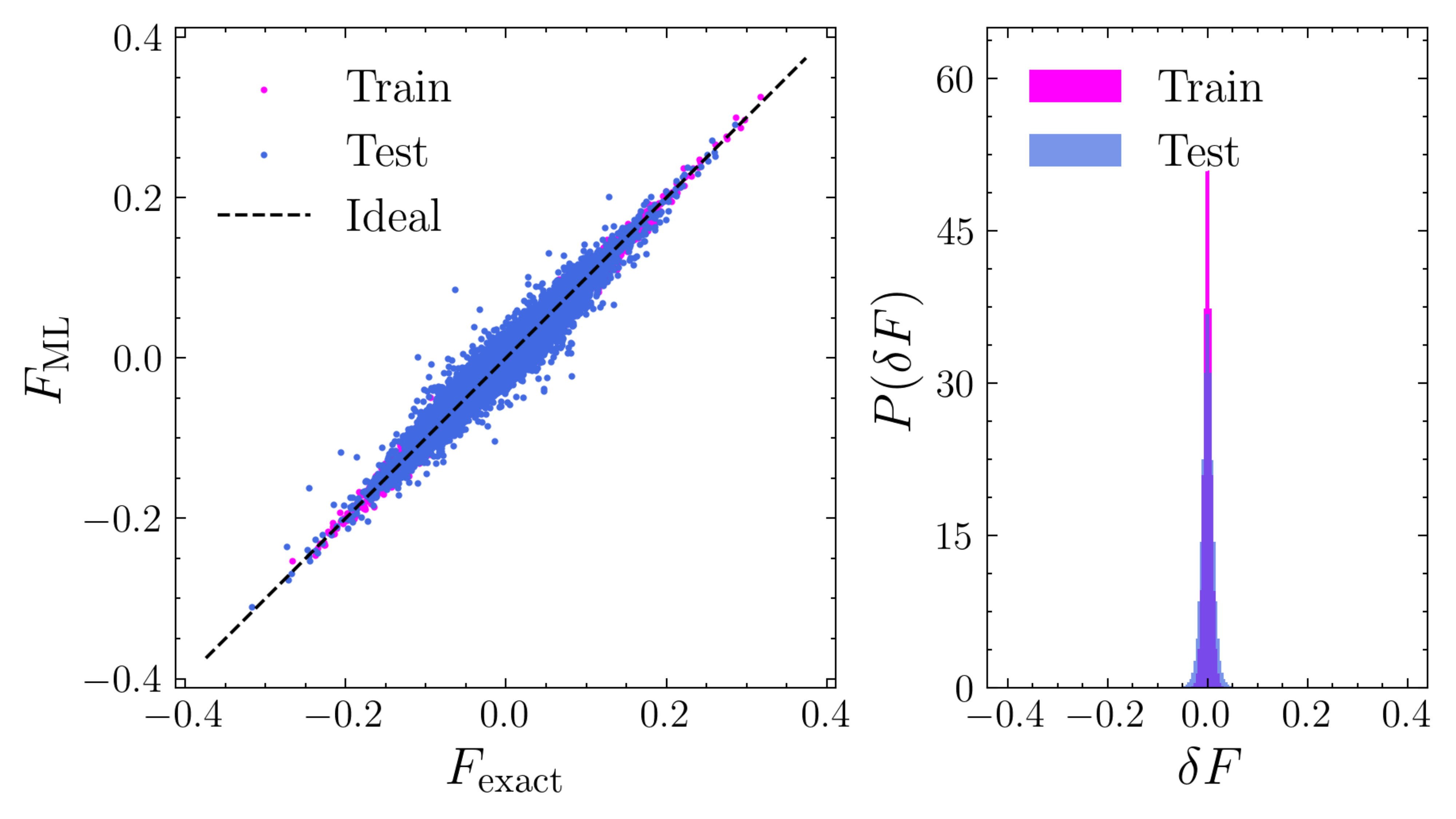}
\caption{
Benchmark of gauge-covariant force prediction for the semiclassical dynamics of the $\mathrm{SU}(2)$ quantum link model. 
Left: parity plot comparing GNN-predicted forces $\mathbf{F}_{ij}^{\mathrm{ML}}$ with exact diagonalization (ED) results $\mathbf{F}_{ij}^{\mathrm{ED}}$ (dashed line: ideal agreement). Right: distribution of prediction errors $\delta F = F^{\mathrm{ML}} - F^{\mathrm{ED}}$ for training (magenta) and test (blue) sets.  The model achieves $\mathrm{MSE} = 1.57 \times 10^{-4}$ and $R^2 = 0.97$, demonstrating accurate learning of fermion-induced, gauge-covariant forces. }
\label{fig:force-benchmark}
\end{figure}

We focus on the semiclassical, adiabatic regime, in which the fermionic sector remains in the instantaneous ground state corresponding to a given gauge configuration $\{U_{ij}\}$. The coupled equations of motion take the form
\begin{align}
	\frac{d}{dt} \mathbf{L}_{ij} &= \mathbf{F}_{ij}[\{U\}], \\
	\frac{d}{dt} U_{ij} &= - i \frac{g^2}{2} \left( \mathbf{L}_{ij} \cdot \boldsymbol{\sigma} \right) U_{ij} + \cdots ,
\end{align}
where $\boldsymbol{\sigma} = (\sigma^1, \sigma^2, \sigma^3)$ denotes the Pauli matrices, providing a basis of the $\mathfrak{su}(2)$ algebra in the fundamental representation. The explicit term corresponds to the canonical left-invariant flow generated by the electric field $\mathbf{L}_{ij}$, while the ellipsis denotes a Casimir-induced phase rotation (proportional to $\mathbf{L}_{ij}^2$) that commutes with all observables and does not affect the force dynamics.

The force acting on each link is determined by the fermionic back-action,
\begin{equation}
	\label{eq:gauge-F}
	\mathbf{F}_{ij} = -  \mathrm{Im}\,\mathrm{Tr}\!\left[ \boldsymbol{\sigma} \, U_{ij} \, \rho_{ij} \right],
\end{equation}
with $\rho_{ij} \equiv \langle \psi_j \psi_i^\dagger \rangle$ the single-particle density matrix evaluated in the instantaneous ground state. This expression makes explicit that the gauge dynamics is entirely mediated by the fermionic response to the background $\{U_{ij}\}$. The dominant computational cost arises from evaluating $\rho_{ij}$ via repeated fermionic diagonalization along the dynamical trajectory.

\begin{figure}[t]
\centering
\includegraphics[width=0.85\columnwidth]{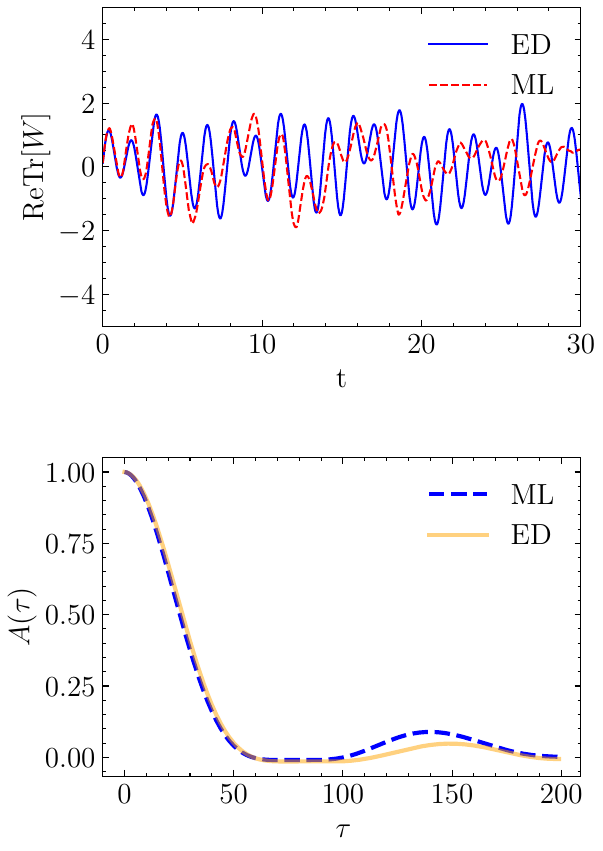}
\caption{Dynamical benchmark of the GNN force-field for semiclassical dynamics of the SU(2) quantum link model. (a)~Time evolution of the gauge-invariant Wilson loop $\mathrm{Re}\,\mathrm{Tr}[W]$ on a representative plaquette, comparing ED forces and ML-predicted forces. The ML dynamics reproduces the short-time behavior accurately, with deviations emerging at longer times due to accumulated prediction errors. (b) Autocorrelation function $A(\tau)$ obtained from $\mathrm{Re}\,\mathrm{Tr}[W]$ averaged over plaquettes and independent trajectories. The ML results are in good agreement with ED across the full time window, with a small deviation at large $\tau$. }
    \label{fig:dyn-benchmark}
\end{figure}

To overcome this bottleneck, we construct a gauge-equivariant GNN to learn the intermediate, gauge-covariant response functional
\begin{equation}
	\mathcal{F}_{\bm\theta}: \, \{U_{ij}\} \mapsto \{\rho_{ij}\},
\end{equation}
directly from data. The physical force is then obtained by inserting the predicted $\rho_{ij}$ into Eq.~(\ref{eq:gauge-F}), ensuring that the learned dynamics preserves the underlying algebraic structure by construction. By replacing repeated fermionic diagonalization with this symmetry-constrained surrogate, the approach enables efficient and scalable simulations of coupled gauge--matter dynamics while maintaining the essential gauge structure.

We benchmark the accuracy of the learned force field in Fig.~\ref{fig:force-benchmark} by comparing the GNN-predicted forces with exact results obtained from fermionic diagonalization. The parity plot exhibits a tight collapse of both training and test data onto the diagonal, indicating that the model faithfully captures the gauge-covariant force across a broad range of dynamical configurations. The corresponding error distribution is sharply peaked around zero, with nearly indistinguishable statistics between training and test sets, demonstrating robust generalization. These results confirm that the gauge-equivariant GNN accurately learns the fermionic response functional underlying Eq.~(\ref{eq:gauge-F}), providing a reliable and symmetry-consistent surrogate for the forces governing semiclassical gauge dynamics.

Having established the accuracy of the GNN in predicting instantaneous forces, we next assess its performance in a fully dynamical setting. To this end, we integrate the trained GNN force field into the semiclassical equations of motion governing the link variables $U_{ij}$ and their conjugate generators $L_{ij}$, and carry out time evolution using the ML-predicted forces. This provides a stringent end-to-end benchmark, as any small local errors in the learned forces can accumulate and propagate nonlinearly during the dynamics.

The results are summarized in Fig.~\ref{fig:dyn-benchmark}. Panel (a) compares the time evolution of a representative gauge-invariant Wilson loop $\mathrm{Re}\,\mathrm{Tr}[W]$ obtained from ML-driven dynamics against that computed using exact-diagonalization (ED) forces. The agreement at short times is excellent, confirming that the learned force field faithfully captures the local structure of the effective energy landscape. At longer times, the trajectories gradually diverge, reflecting the expected accumulation of small prediction errors in a nonlinear dynamical system. Importantly, this trajectory-level sensitivity does not preclude accurate statistical predictions. As shown in panel (b), the autocorrelation function $A(\tau)$---averaged over plaquettes and independent initial conditions---remains in close agreement between ML and ED across the entire time window. The weak feature at large $\tau$ is reproduced qualitatively by both approaches and likely reflects finite-size effects or residual dynamical recurrences. Together, these results demonstrate that the GNN force field, when coupled to the semiclassical equations of motion, provides an accurate and robust description of the emergent dynamical behavior at the level of physically relevant observables.

\section{Discussion}

In this work, we introduced a general gauge-equivariant graph neural network framework for lattice gauge systems, in which local non-Abelian symmetry is enforced exactly at the architectural level through matrix-valued message passing. By operating directly on gauge-covariant degrees of freedom and preserving transformation laws throughout, the approach eliminates the need for explicit construction of gauge-invariant features while retaining full structural information. We demonstrated its effectiveness across three representative regimes. In pure gauge theory, the network reconstructs both global and local gauge-invariant observables directly from covariant inputs. In gauge–matter systems, it captures fermion-mediated, intrinsically nonlocal correlations with high accuracy despite its local message-passing structure. In dynamical settings, it provides symmetry-consistent force fields that faithfully reproduce the statistical properties of semiclassical gauge dynamics. Together, these results establish gauge-equivariant message passing as a unified and scalable paradigm for learning both static and dynamical properties of non-Abelian lattice gauge systems.

Looking forward, the present framework opens several promising directions at the interface of machine learning and lattice gauge theory. A natural application lies in accelerating lattice QCD simulations. By embedding gauge-equivariant neural surrogates into Monte Carlo workflows—such as learned effective actions, proposal generators, or force models—one may construct more efficient sampling schemes while preserving exactness through standard accept–reject steps~\cite{albergo2019,kanwar2020,abbott2022,nagai2023,nagai2024}. This direction is particularly compelling in gauge–matter settings, where fermionic contributions render conventional updates computationally expensive. Beyond sampling, the same architecture may serve as a physics-informed accelerator for fermionic solvers, for example by providing symmetry-consistent preconditioners or surrogate representations of Dirac operators, potentially alleviating the dominant cost associated with repeated inversions.

From the perspective of condensed matter and quantum simulation, the same framework naturally extends to variational representations of quantum states in gauge-constrained many-body systems~\cite{luo2021,luo2023,cheng2025}. Existing neural quantum state constructions in this context have thus far been predominantly developed for Abelian lattice gauge theories, where gauge constraints can be incorporated more directly into the variational ansatz. By interpreting gauge-invariant outputs as amplitudes of a many-body wavefunction, the equivariant message-passing backbone defines a scalable neural ansatz that enforces local gauge constraints by construction. Combined with variational Monte Carlo, this enables compact and systematically improvable descriptions of ground states in both Abelian and non-Abelian models. Furthermore, integration with time-dependent variational principles provides a route to real-time simulations of nonequilibrium phenomena—including quenches, transport, and confinement dynamics—within a symmetry-preserving variational manifold. These directions are particularly relevant for quantum link models and other finite-dimensional gauge systems realizable in ultracold-atom and programmable quantum platforms, where symmetry-consistent, data-driven approaches can directly interface with experimental platforms.

Enforcing local gauge symmetry at the architectural level thus reshapes how machine learning interfaces with lattice gauge theory. Rather than approximating gauge-invariant observables through engineered features, the present framework learns directly from gauge-covariant degrees of freedom and organizes them into physical predictions. This enables a unified treatment of static observables, fermion-mediated interactions, and dynamical evolution, with particular impact in non-Abelian settings where locality and symmetry are deeply intertwined. More generally, our results point toward a paradigm in which symmetry-preserving architectures function as structure-aware extensions of first-principles theory, providing a scalable route to modeling complex quantum systems beyond the reach of brute-force approaches.

\par\vspace*{6pt}
\begin{acknowledgments}
This work was supported by the U.S. Department of Energy, Office of Basic Energy Sciences, under Contract No.~DE-SC0020330. Y.L. acknowledges support from the U.S. Department of Energy under Contract No.~DE-SC0024644. The authors thank M.~Engelhardt and S.~Liuti for insightful discussions on lattice QCD simulations and phenomenological analyses. The authors also acknowledge Research Computing at the University of Virginia for providing computational resources and technical support that contributed to the results reported in this work.
\end{acknowledgments}

\bibliography{ref.bib}

\end{document}